\RequirePackage{fix-cm}
\documentclass[smallextended]{svjour3}       
\smartqed  
\usepackage{graphicx}
\usepackage[T1]{fontenc}

\usepackage{amsmath}
\usepackage{amsfonts}
\usepackage{mathtools}
\usepackage[mathscr]{euscript}
\usepackage[pdftex]{hyperref}

\usepackage{comment}
\usepackage{graphicx}
\usepackage[export]{adjustbox}
\usepackage[caption=false]{subfig}
\usepackage{algorithm}
\usepackage[noend]{algpseudocode}
\usepackage{amsthm}
\usepackage{rotating}
\usepackage[dvipsnames]{xcolor}
\usepackage{bm}
\usepackage{multirow}

\usepackage{enumitem}

\def\X{{\mathbf{X}}}
\def\Y{{\mathbf{Y}}}
\def\x{{\mathbf{x}}}

\def\z{{\mathbf{z}}}
\def\y{{\mathbf{y}}}

\def\z{{\mathbf{z}}} 
\def\E{{\mathbb E}}
\def\V{{\mathbb V}}

\renewcommand{\vec}{\mathbf}

\usepackage{natbib}

\begin{document}

\title{Spectral smooth tests for goodness-of-fit
}


\author{Victor Cândido Reis         \and
        Rafael Izbicki 
}


\institute{Victor Cândido Reis \\ Federal University of São Carlos \at
              Rod. Washington Luís, s/n - Monjolinho, São Carlos - SP, 13565-905\\
              \email{Rafael Izbicki \\ Federal University of São Carlos \at
              Rod. Washington Luís, s/n - Monjolinho, São Carlos - SP, 13565-905\\
              \email{rafaelizbicki@gmail.com} }           
           \and
          Rafael Izbicki \\ Federal University of São Carlos \at
              Rod. Washington Luís, s/n - Monjolinho, São Carlos - SP, 13565-905\\
              \email{rafaelizbicki@gmail.com} 
}

\date{Received: date / Accepted: date}

\maketitle

\begin{abstract}
Goodness-of-fit tests are crucial tools for assessing the validity of statistical models. In this paper, we introduce a novel approach, the Spectral Smooth Test (SST), that generalizes Neyman's smooth test to high-dimensional data settings. While conventional goodness-of-fit tests for univariate data are well-established, extending them to high dimensions, such as images, trajectories, and SNPs, poses significant challenges. Our proposed SST leverages spectral bases, which adapt naturally to the geometry of feature spaces, to model multivariate distributions. Unlike traditional orthogonal bases, these spectral bases are tailored to the data distribution, enabling more effective function modeling. The SST framework offers a principled way to estimate the underlying model, thereby providing actionable insights even when the null hypothesis is rejected. We present experimental results demonstrating the robustness of SST across various tuning parameter choices and compare its performance against other goodness-of-fit tests. Furthermore, we apply SST to the MNIST dataset as a real-world example, showcasing its effectiveness in high-dimensional scenarios. 
\end{abstract}

\section{Introduction}
\label{sec:intro}

Hypothesis tests are widely used by scientists. In particular, goodness-of-fit tests are  important because they enable practitioners to assess statistical models, and thus are a key tool for creating accurate descriptions of the world. 
In its simplest formulation, the goal of a goodness-of-fit test is to test the null hypothesis that an i.i.d. sample, $\X_1,\ldots,\X_n$, comes from a postulated distribution $F_0$.

Although there exists a large body of work that develops   goodness-of-fit tests for univariate settings \citep{mann1947test,kolmogorov1933sulla,smirnov1948table,algeri2020exhaustive}, it was only recently that nonparametric  tests designed for high dimensional data have gained interest   \citep{gretton2012kernel,ramdas2017wasserstein,pfister2018kernel,kim2018robust,kim2019global,inacio2021distance,polo2022unified,mcneely2023detecting}.  This is because such tests are essential to deal with complex objects such as images, trajectories and  SNPs.
These tests typically have good statistical power, but they are often hard to interpret: The output of such tests are just
reject/don't reject decisions, that do not offer insights on \emph{how} a given model could be improved.

In the univariate context, 
an alternative approach to hypothesis testing is  Neyman's smooth test  \citep{neyman1937}, which has a key advantage: when the null hypothesis is rejected, it automatically yields an estimate for the correct model \citep{shalizi2013advanced}. This is because Neyman's approach builds on a density estimator of the distribution of $\X$. More specifically, in the special case where $\X \in (0,1)$ and $H_0:X \sim U(0,1)$,
the key idea of Neyman's  test is to represent the density of $\X$ as $$f_\theta(\x) \propto \exp{\left(\sum_{i=1}^\infty \theta_i \phi_i(\x)\right)},$$
 where  $(\phi_i)_{i \geq 0}$ is an orthonormal basis  with $\phi_0(\x)\equiv 1$
  (such as the Fourier basis). Because the distribution of $\X$ is uniform   (that is, $f_\theta(\x)=1$) if, and only if, $ \theta_1=\theta_2=\ldots=0,$
the idea is then to test $H_0$
 by first estimating $\theta_i$'s and then computing
the test statistic 
$T=\sum_{i=1}^I (\widehat{\theta}_i)^2,$
where $I$ is a tuning parameter that needs to be chosen. The test rejects $H_0$ if $T$ is large enough. If the null hypothesis is rejected, 
$\widehat f(\x) \propto \exp{\left(\sum_{i=1}^I \widehat \theta_i \phi_i(\x)\right)}$ can be used as an alternative model. The estimate $\widehat f(\x)$ can also be used to quantify \emph{how} $f$ differs from the null model (see \citealt{algeri2020exhaustive} for ways in which this insights can be used). 

Several extensions of Neyman's original proposal exist. For instance,   \citet{thomas1979neyman} work with cases where the null hypothesis is composite, and
\citet{xiao2004smooth} propose a version of the test that can be used for two-sample comparisons.  See  \citet{rayner1990smooth,ghosh2002neyman,algeri2020exhaustive}  and references therein for other extensions. To the best of our knowledge, none of such extensions is able to deal with high-dimensional data.
In this work, we propose a generalization that smooth tests that can be used for high dimensional data, the \emph{Spectral Smooth Test} (SST). This setting is a challenge for the original formulation because
it is computationally expensive to extend
Fourier and other data-independent unidimensional bases to higher dimensions. Indeed, this extension is typically done by using tensor products \citep{efromovich2008nonparametric}, which quickly become computationally intractable even for as few as five
variables. In order to overcome this issue, we instead use  \emph{spectral bases}
 \citep{LeeIzbickiReg,IzbickiLeeCDE} to model multivariate distributions. 
 Spectral bases depend on the distribution of the data and naturally adapt to the geometry of the the feature space, and thus they can do a better job in modeling functions defined over the data domain. However, because spectral basis are orthogonal with respect to the data distribution (rather than to the Lebesgue measure), Neyman's approach needs to be adapted.

Section \ref{sec:method}  introduces our approach, the \emph{spectral smooth test}.
 Section \ref{sec:experiments}  shows experiments that evaluate its robustness to the choice of tuning parameters. It also presents comparisons to other goodness-of-fit tests. 
 Section \ref{sec:app} presents an application of SST to the MNIST dataset \citep{deng2012mnist}.
 Section \ref{sec:final} concludes the paper.

\section{Methodology}
\label{sec:method}

\subsection{Diffusion basis}

We start by reviewing the diffusion basis
\cite{lee2010spectral}, the basis functions we use in SST.
Let $F_0$ be a probability measure over $\mathcal{X} \subseteq \mathbb{R}^d$ and   $k(\z,\y)$ be a Mercer Kernel defined over
$\mathbb{R}^d \times \mathbb{R}^d$.
The diffusion operator \cite{lee2010spectral} is defined as 
$$A(h)(\z)=\int_{\mathcal{X}} a(\z,\y)h(\y)dF_0(\y),$$
where
$$a(\z,\y)=\dfrac{k(\z,\y)}{p(\z)}$$
and
$p(\z)=\int_{\mathcal{X}} k(\z,\y)dF_0(\y).$
The diffusion basis $\{\psi_i(\z)\}_{i \ge 0}$ is composed of the eigenfunctions of this operator. We sort them according to their  eigenvalues, $1 = \lambda_0 \ge \lambda_1 \ge \cdots \ge 0$. These  eigenfunctions are orthonormal with respect to the measure $S(\mathcal{C}) \propto \int_{\mathcal{C}} p(\z)dP(\z)$, that is,
$$\int_{\mathcal{X}} \psi_i(\z)\psi_j(\z)dS(\z)=\delta_{i,j}.$$

The diffusion basis $\{\psi_i(\z)\}_{i \ge 0}$ adapts to the intrinsic geometry of the data, forming an efficient Fourier-like
orthogonal basis for smooth functions on the data \citep{LeeIzbickiReg}. More precisely, they are concentrated around high-density regions,  where
lower-order terms are smoother than higher-order terms.
Thus, this basis  can efficiently model smooth functions in $\mathcal{X}$, meaning that only the first few terms of the series are needed. Moreover, they have the property that
$\psi_0(\z) \equiv 1$.
As we will see in the next section, this property is key to develop our test statistic.

\subsection{Rewriting the null hypothesis}

We now show how diffusion basis can be used to model the distribution of the data and  how it can be used to test goodness-of-fit.
Let $\X_1,\cdots,\X_n \in \mathbb{R}^d$ be independent and identically distributed to
$F$. Our goal is to test the null hypothesis
$H_0: F=F_0$. We assume that $F_0$ dominates $F$. 
The null hypothesis is therefore equivalent to
$$H_0: h(\x) \equiv 1,$$
where 
$$h(\x):=\dfrac{dF}{dF_0}$$ 
is the Radon–Nikodym  density of $F$
with respect to $F_0$. 

Next, let $\{\psi_i(\x)\}_{i \ge 0}$  be the  diffusion basis with respect to $F_0$. Notice that as long as $h \in \mathcal{L}^2(\mathcal{X},S)$, this density admits the representation
\begin{align}
\label{eq:radon}
h(\x)= \sum_{i \ge 0} \theta_i \psi_i(\x),
\end{align}
where
\begin{align}
\label{eq:projection}
\theta_i&=\int_{\mathcal{X}} h(\x)\psi_i(\x)dS(\x) = \int_{\mathcal{X}} \dfrac{dF}{dF_0}(\x)\psi_i(\x)s(\x)dF_0(\x) \notag \\
&=\int_{\mathcal{X}} \psi_i(\x)s(\x)dF(\x)
=\mathbb{E}_{F}[\psi_i(\X)s(\X)].    
\end{align}
Now, Equation \ref{eq:projection} implies that the projection of 
any density into $\psi_0$ is 1, and thus $\theta_0=1$. Moreover, the projection of
the function $g(\x)\equiv 1$ on the basis  
is given by 
and $\theta_i=0$, $i>0$. Thus, the null hypothesis may be written as 
\begin{align}
\label{eq:null}
H_0: \theta_0=1  \text{ and }\theta_i=0 \text{ for every } i>0.    
\end{align}
It follows that we can test the null hypothesis by testing whether the expansion coefficients are zero.

\subsection{Estimating the expansion coefficient}

The expansion coefficients in   Equation \ref{eq:null} are parameters. In this section we discuss how to estimate them, and in the next section we show how to use such estimates to test $H_0$.

First we will estimate the diffusion basis by using the approach from \citet{Lee}.
We start by generating an i.i.d. sample from $F_0$ with size $m$, $\Y_1,\ldots,\Y_m$. Next, let $\mathbb{K}$ be the $m \times m$ Gram matrix whose $(i,j)$ entrance is given by $k(\y_i,\y_j)$. For simplicity, we use the gaussian kernel, 

\begin{equation}
\label{kernel}
    k(\y_i,\y_j)=\exp{\left(-\frac{d^2(\y_i,\y_j)}{\epsilon}\right)},
\end{equation}
although other kernels could be used as well. Next, we renormalize each row of the matrix according to $\mathbb{A}=\mathbb{D}^{-1}\mathbb{K},$ where $\mathbb{D}$ is the a diagonal matrix where the element $(i,i)$ is $\sum_{l=1}^m K(\y_i,\y_l)$.
Finally, we compute the  $I$ first right eigenvectors of $\mathbb{A}$ following Appendix \ref{sec:algorithms}, $\widehat \psi_i(\y_1),\ldots,\widehat \psi_i(\y_m),$ $i=0,\ldots,I$ (we discuss the choice of $I$ in Section \ref{sec:test_statistic}).
Our estimate of $\psi_i(\x)$ is given by the Nystr\"{o}m extension of such eigenvectors:
\begin{equation*} 
\widehat{\psi}_{i}(\x)  =\frac{1}{\lambda_{i}} \sum_{j=1}^{m}  \frac{k(\x,\y_j)}{\sum_{l=1}^{n} k(\x,\y_l)} \widehat{\psi}_{i}(\y_j),
\end{equation*}
where $\lambda_{i}$ are the eigenvalues of $\mathbb{A}$.

Equation \ref{eq:projection} suggests  the following estimator to the expansion coefficients:
\begin{align*}
    \widehat{\theta}_i=\frac{1}{n}\sum_{k=1}^n  \widehat{\psi}_i(\x_k)\vec{s}(\x_k),
\end{align*}
where
$$\vec{s}(\x_k)=\dfrac{\dfrac{1}{m}\sum_{j=1}^m k(\x_k,\y_j)}{\dfrac{1}{m}\sum_{i=1}^m \dfrac{1}{m} \sum_{l=1}^m k_{\epsilon}(\y_i,\y_l)}=\dfrac{\sum_{j=1}^m k(\x_k,\y_j)}{\dfrac{1}{m}\sum_{i=1}^m \sum_{l=1}^m k(\y_i,\y_l)}.$$

\subsection{Test statistic}
\label{sec:test_statistic}

Inspired by Neyman's test statistic, we can test  $H_0$ by  using the  statistic 
\begin{align*}
T_{\lambda}=(\widehat{\theta}_0-1)^2+\sum_{i=1}^{I} \widehat{\theta}_{i}^2,    
\end{align*}
where $\lambda$ denotes the dependence of the test statistic on tuning parameters associated to the kernel (such as the bandwidth $\epsilon$ in the case of a Gaussian kernel) as well as on the cutoff $I$. The p-value against $H_0$, $p_\lambda:=P_0(T_{\lambda}^*>t_{\lambda}^{*(obs)})$, can be estimated via Monte Carlo sampling at $F_0$. Let $\widehat p_\lambda$ be such estimate.

Instead of relying solely on a single choice $\lambda$, we combine several of these choices into a new test statistic. As we will show in Section \ref{subsec:robustnees}, this will lead to robust results. 
In order to do so, let $\E_0(T_{\lambda})$ and $\V_0(T_{\lambda})$ denote the mean and variance of  $T_{\lambda}$ under the null hypothesis, and let $\widehat \E_0(T_{\lambda})$ and $\widehat \V_0(T_{\lambda})$ be estimates of these quantities obtained by Monte Carlo sampling at $F_0$. First, we define  $T_{\lambda}^*$, a re-scaled version of $T_{\lambda}$, according to:
\begin{align*}
   T^*_{\lambda}= \frac{T_{\lambda}-\widehat \E_0(T_{\lambda})}{\sqrt{\widehat  \V_0(T_{\lambda})}}.
\end{align*} 
The SST combined test statistic is given by
\begin{align*}
    T_{SST}=\sum_{\lambda \in \Lambda} (1-\widehat p_\lambda)|T_\lambda^*|,
\end{align*}
where $\Lambda$ denotes a  set of predefined choices
for $\lambda$. The idea is that if $T_\lambda$ is close to what is expected under the null, then the p-value associated to $\lambda$ should not receive a large weight relative to the other ones. Thus, the combined statistic weights p-values according to $|T^*_{\lambda}|$.

Again, p-values for $T_{SST}$ can be approximated by  Monte Carlo sampling at  $F_0$.

\section{Experiments}
\label{sec:experiments}

Next, we show  simulation studies to evaluate the robustness and power of our test. 
In Section \ref{subsec:robustnees}, we evaluate the sensitivity of the tuning parameters $\lambda$ on the test statistic $T_{\lambda}$, as well as the robustness of the combined statistic $T_{SST}$. In Section \ref{subsec:capariontoother} we compare SST with other procedures on both univariate and multivariate examples.

\vspace{0.5cm}

All simulations studies share the following characteristics:
\begin{enumerate}
    \item The sample size is taken to be $n=50$.
    \item The diffusion basis was estimated using $10^4$ sample points from  $F_0$.
    \item  $2\times10^4$ Monte Carlo samples from $F_0$  were used to estimate $p_\lambda$, $\E[T_\lambda]$, and $\V[T_\lambda]$ (for all $\lambda$'s)
    \item $10^4$ Monte Carlo samples from $F_0$  were used to estimate the null distribution of $T_{SST}$.

    \item We used all thresholds in $I=\{1, \cdots, 10\}$ and 10
bandwidths. The bandwidths were chosen computing quantiles
of the square distances between $10^4$ samples from $F_0$. We used the following quantiles:
\begin{align*}
    \{&q_{16.667\%},
q_{33.333\%},q_{50\%},q_{66.667\%},q_{0.833\%}, 2 \times q_{0.833\%},3 \times q_{0.833\%},\\
&4 \times q_{0.833\%},5 \times q_{0.833\%},
6 \times q_{0.833\%}\}.
\end{align*}

\item The significance level was set to $\alpha = 0.05$, and  $1000$ repetitions of each setting were used to estimate the power function.
	\item The Gaussian Kernel (Equation \ref{kernel}) was used.
\end{enumerate}

We compare the methods using all scenarios described in  \cite{Ceregatti2020}, but adapted to the goodness-of-fit setting. They are the following:
\begin{enumerate}[label=(\alph*)]
	\item Normal mean shift: $F_0 \sim N(0,1)$ and $\textbf{X} \sim N(\theta,1)$, $\theta =0,\cdots, 0.7$
	\vspace{3mm}
	
	\item Normal variance shift: $F_0 \sim N(0,1)$ and $\textbf{X} \sim N(0, \theta)$, $\theta =1,\cdots, 2.5$
	\vspace{3mm}
	
    \item LogNormal mean shift:  $F_0 \sim LogNormal(0,1)$ and $\textbf{X} \sim LogNormal(\theta, 1)$, $\theta = 0, \cdots, 1$
    \vspace{3mm}
    \item LogNormal variance shift: $F_0 \sim LogNormal(0,1)$ and $\textbf{X} \sim LogNormal(0, \theta)$, $\theta = 1, \cdots, 2.5$
\vspace{3mm}

	\item Beta symmetry: $F_0 \sim Beta(1,1)$ and $\textbf{X} \sim Beta(\theta,\theta)$, $\theta =1,\cdots, 5$
	\vspace{3mm}
	
	\item Gamma Shape: $F_0 \sim Gamma(3,2)$ and $\textbf{X} \sim Gamma(\theta,2)$, $\theta =3,\cdots, 4.5$
	\vspace{3mm}
	
	\item Normal mixtures: $F_0 \sim N(0,1)$ and $\textbf{X} \sim \frac{1}{2}N(-\theta,1)+\frac{1}{2}N(\theta,1)$, $\theta = 0, \cdots, 2$
	\vspace{3mm}
	
	\item Fat tails: $F_0 \sim N(0,1)$ and $\textbf{X} \sim t(\theta^{-1})$, $\theta = 10^{-3}, \cdots, 1$.
\end{enumerate}

All analyses were performed in R \citep{team2013r}.

\subsection{Robustness to tuning parameters}
\label{subsec:robustnees}

\begin{figure}[H]
\centering
\subfloat[$N(\mu=\theta,\sigma=1)$]{\includegraphics[scale=.2]{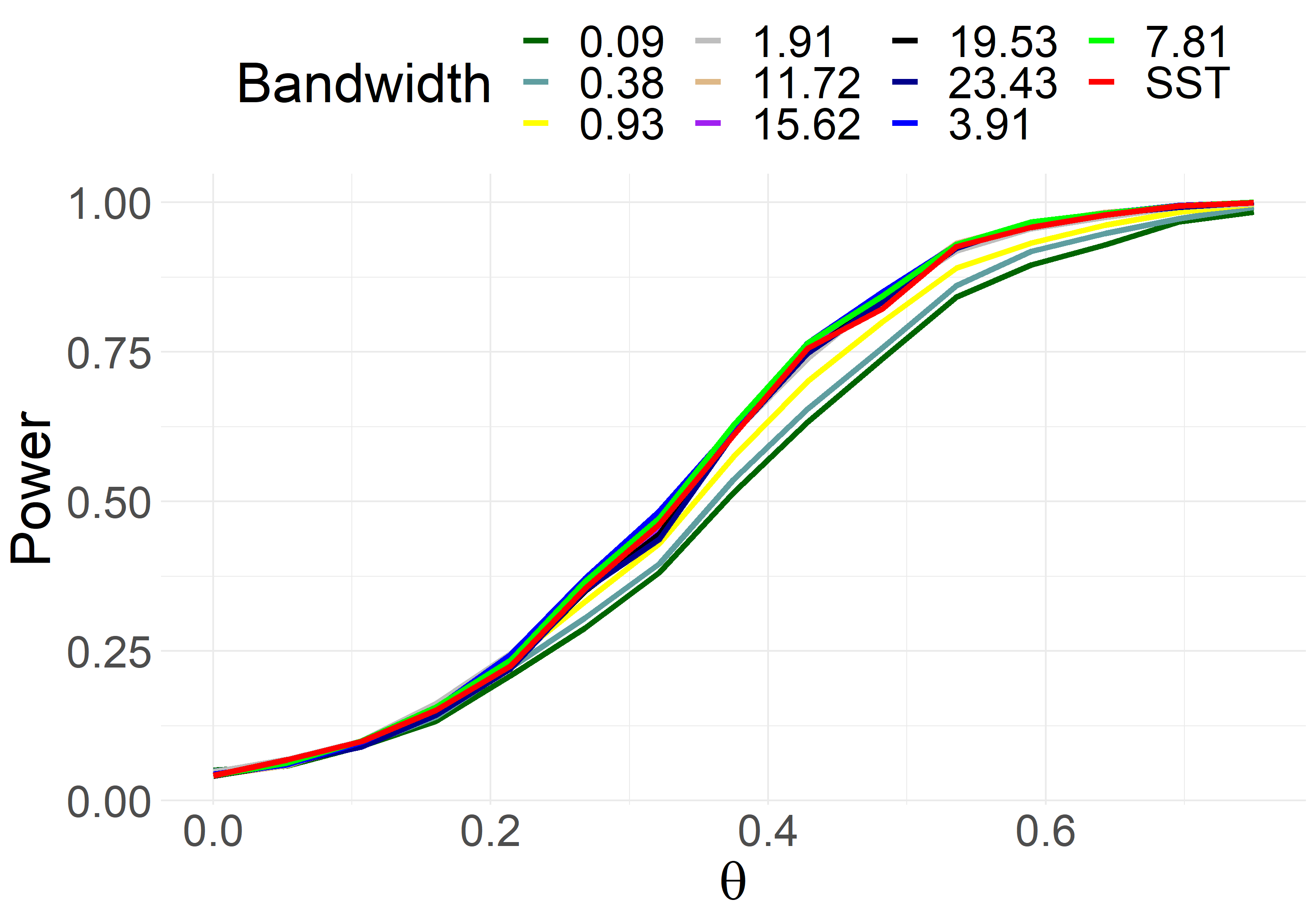}
}
\subfloat[$N(\mu=0,\sigma=\theta)$]{\includegraphics[scale=.2]{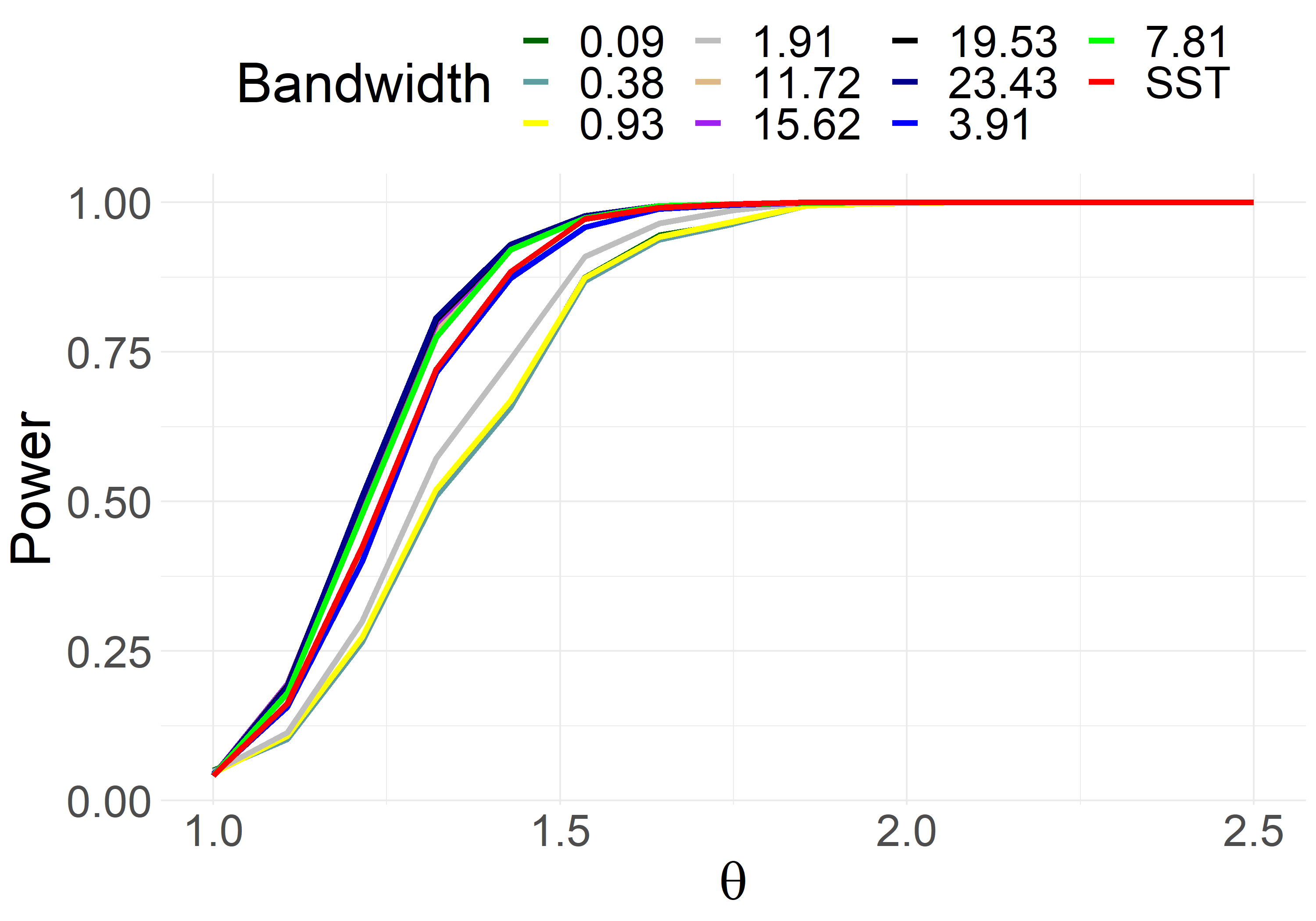}
}

\caption{Robustness to tuning parameters. While the power of the test statistic $T_\lambda$ is heavily affected by the choice of the bandwidth, the combined test statistic (SST) is able to achieve robust results.} 
\label{fig:robustness}
\end{figure}

We evaluate the power $T_\lambda$ as a function of the choice of the  kernel bandwidth in settings  (a) and (b). Figure \ref{fig:robustness} shows that the power of $T_\lambda$ is very sensitive to such choice. At the same time, the combined approach (SST) leads to a test with power close to the best choice of $\lambda$, removing the impact of bad choices of this tuning parameter. 

\subsection{Comparison to other methods}
\label{subsec:capariontoother}

\begin{figure}[H]
\centering
\subfloat[$N(\mu=\theta,\sigma=1)$]{\includegraphics[scale=.2]{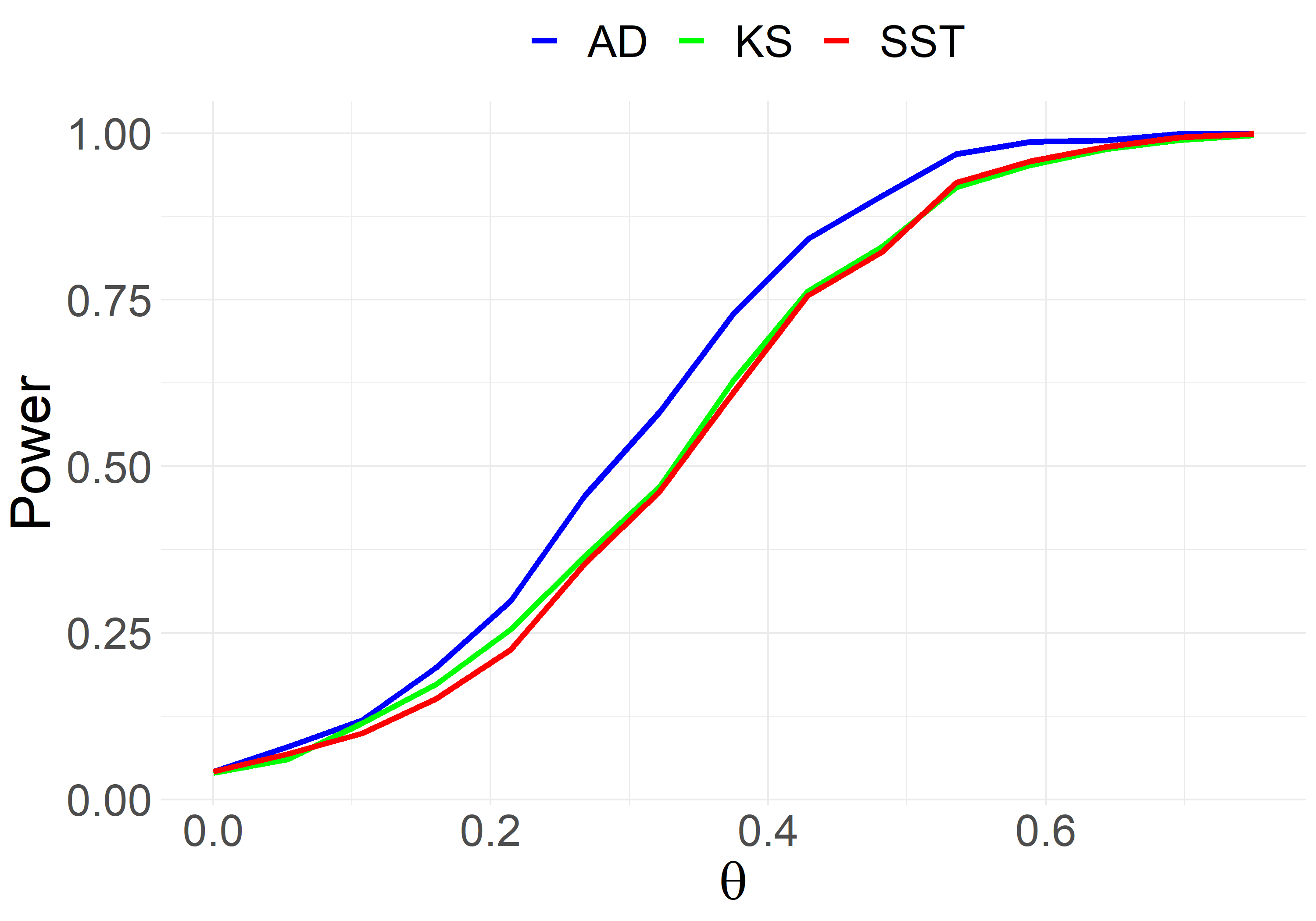}
}
\subfloat[$N(\mu=0,\sigma=\theta)$]{\includegraphics[scale=.2]{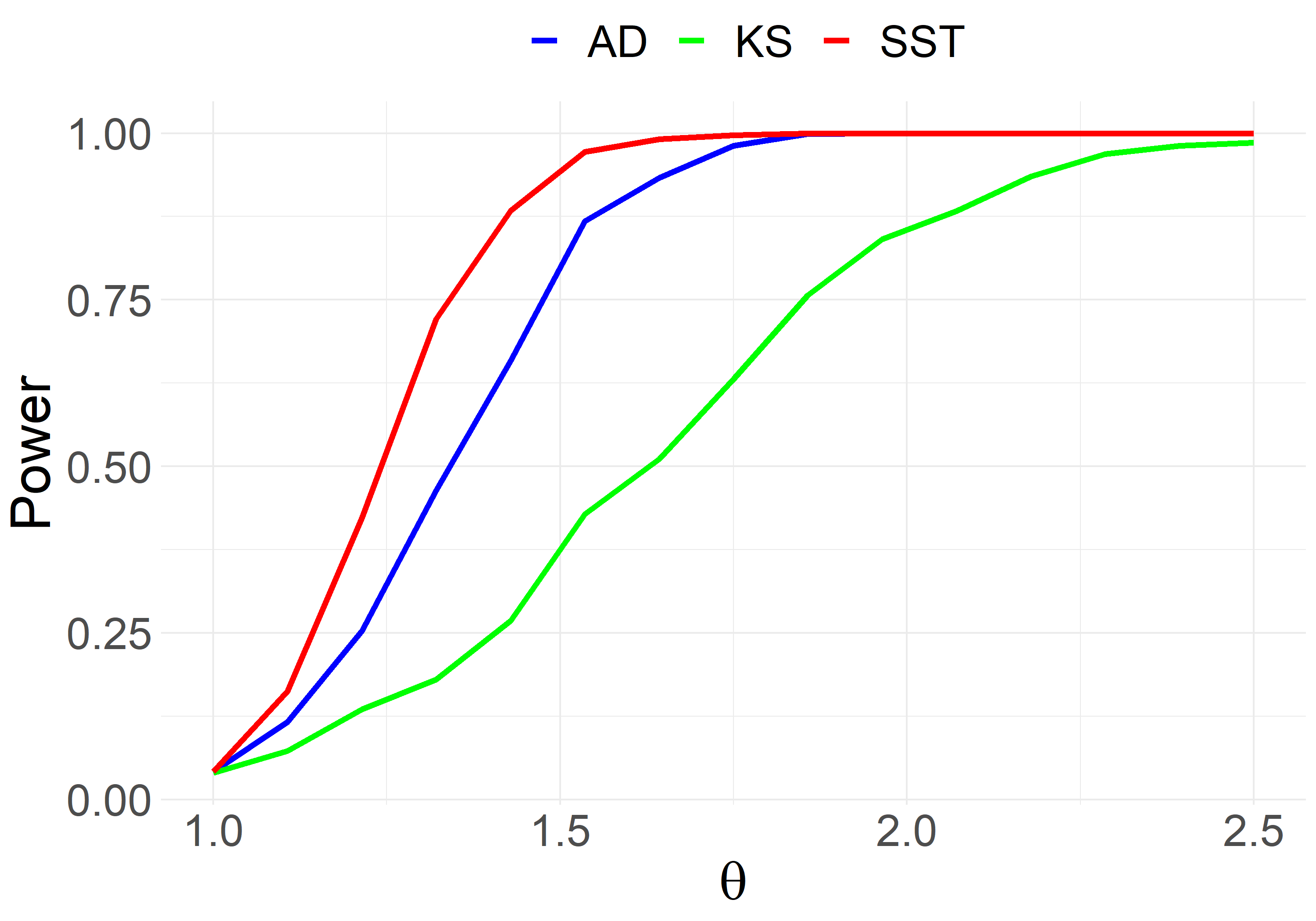}
}\\

\subfloat[$LogNormal(\mu=\theta,\sigma=1)$]{\includegraphics[scale=.2]{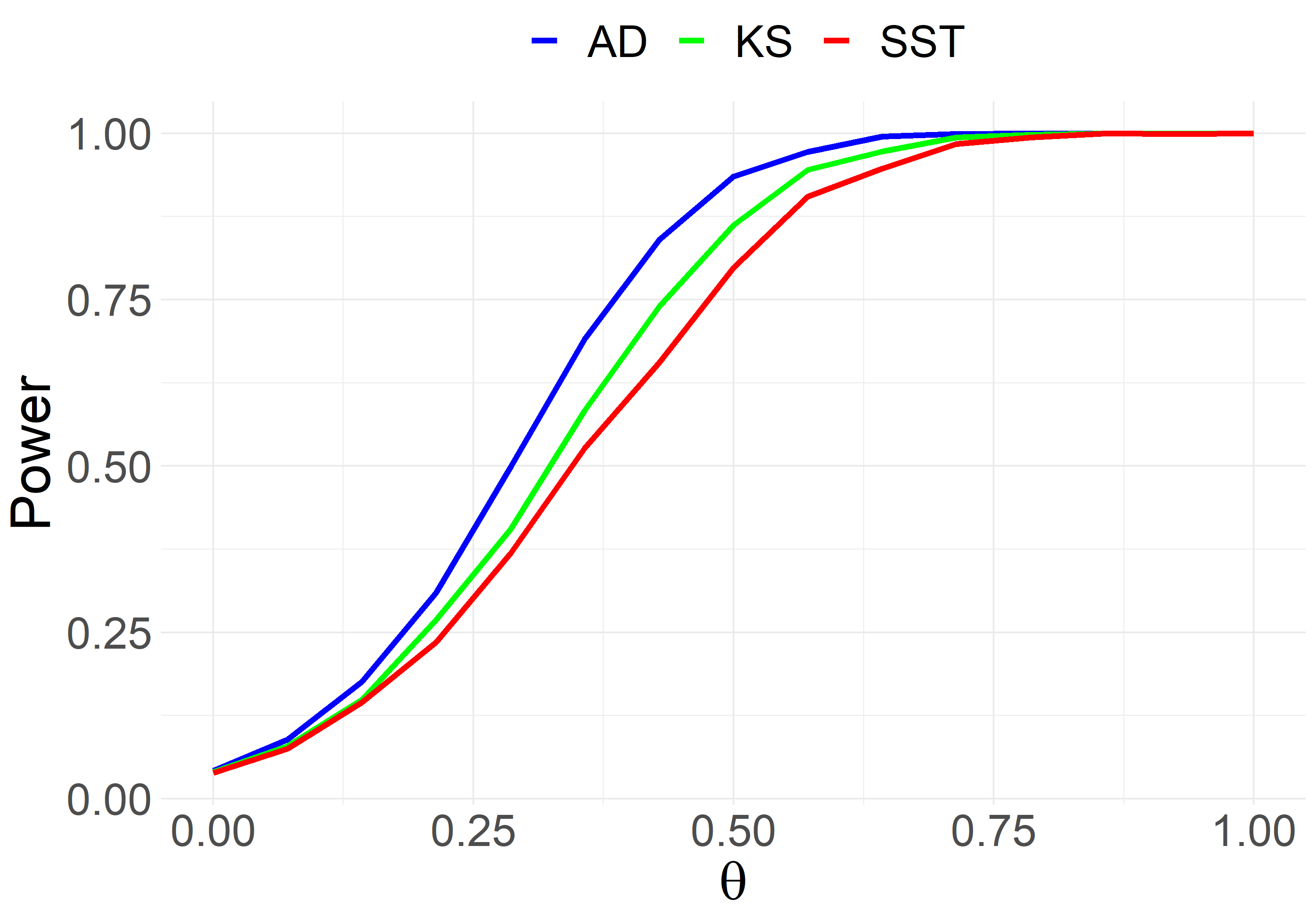}
}
\subfloat[$LogNormal(\mu=0,\sigma=\theta)$]{\includegraphics[scale=.2]{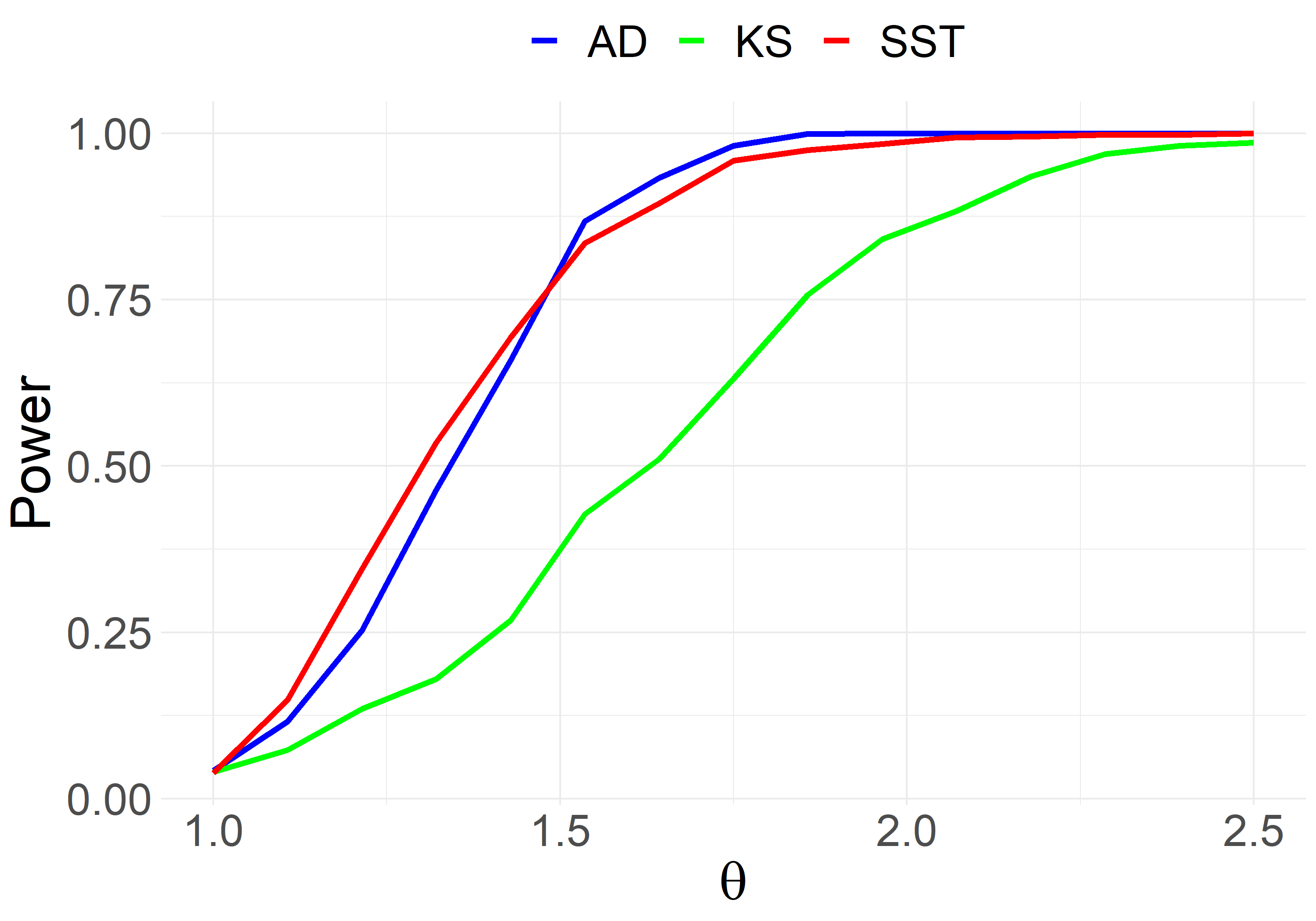}
}\\

\subfloat[$Beta(\alpha=\theta,\beta=\theta)$]{\includegraphics[scale=.2]{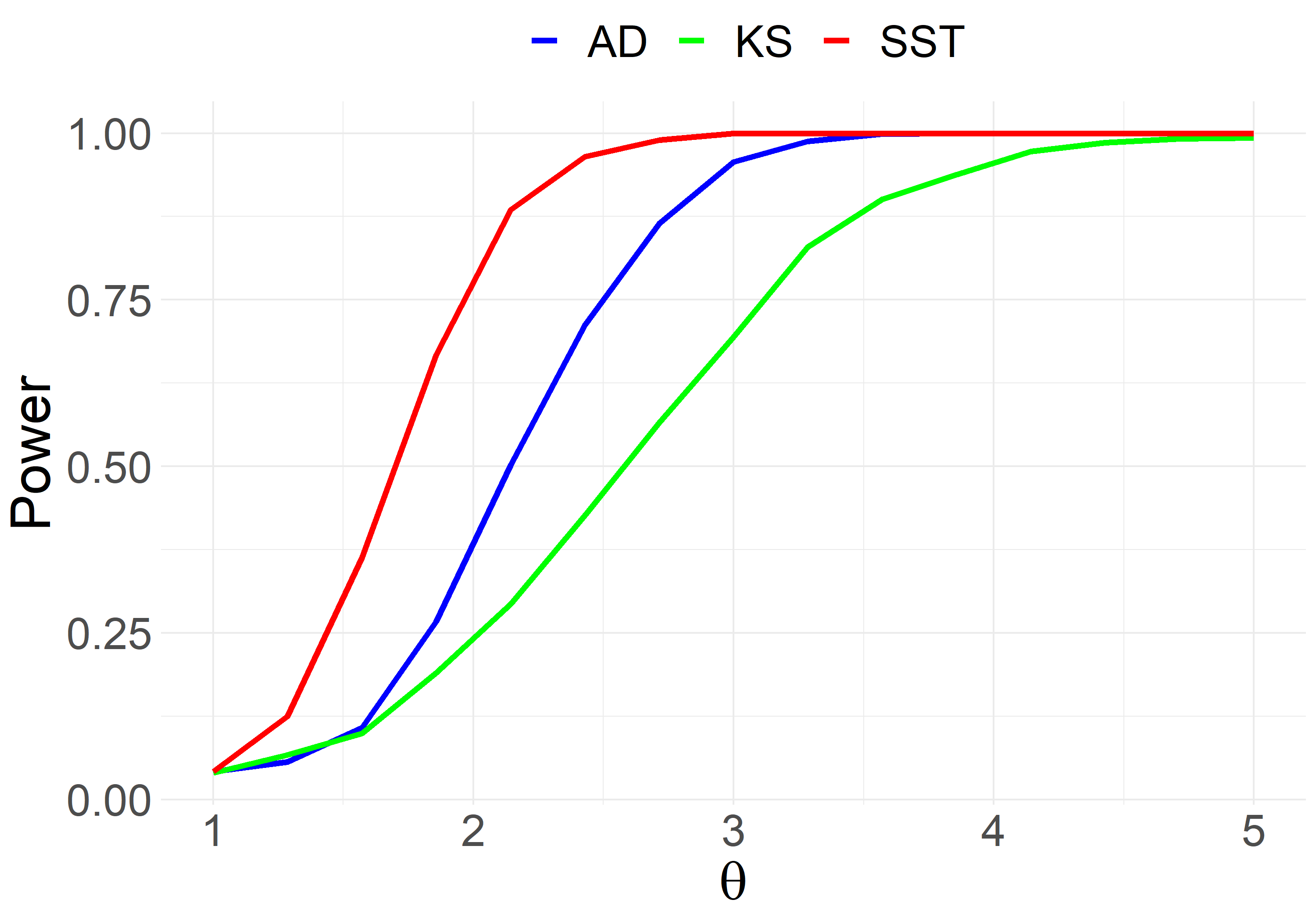}
}
\subfloat[$Gamma(\alpha=\theta,\beta=2)$]{\includegraphics[scale=.2]{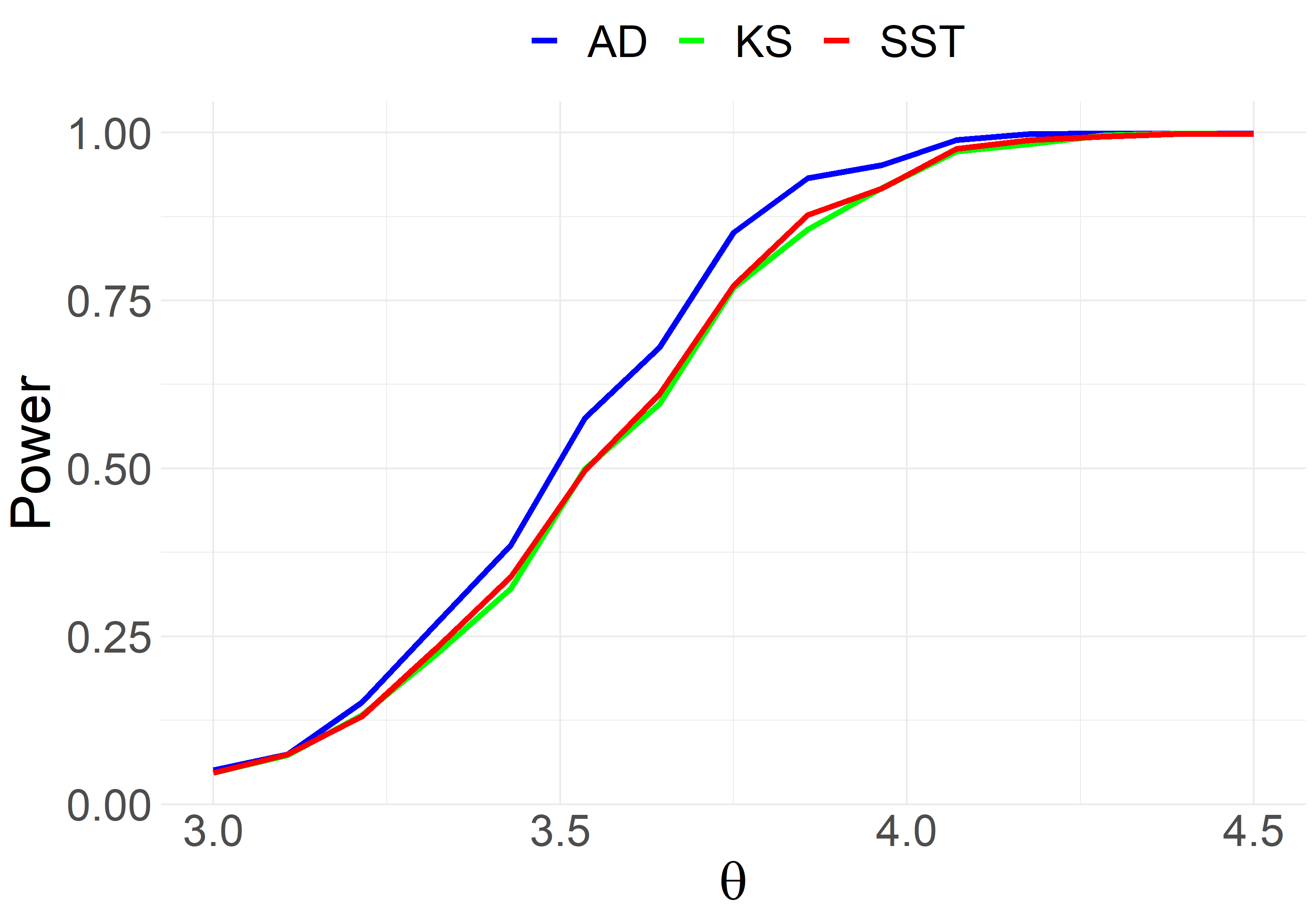}
}\\

\subfloat[$N(\mu=\pm \theta,\sigma=1)$]{\includegraphics[scale=.2]{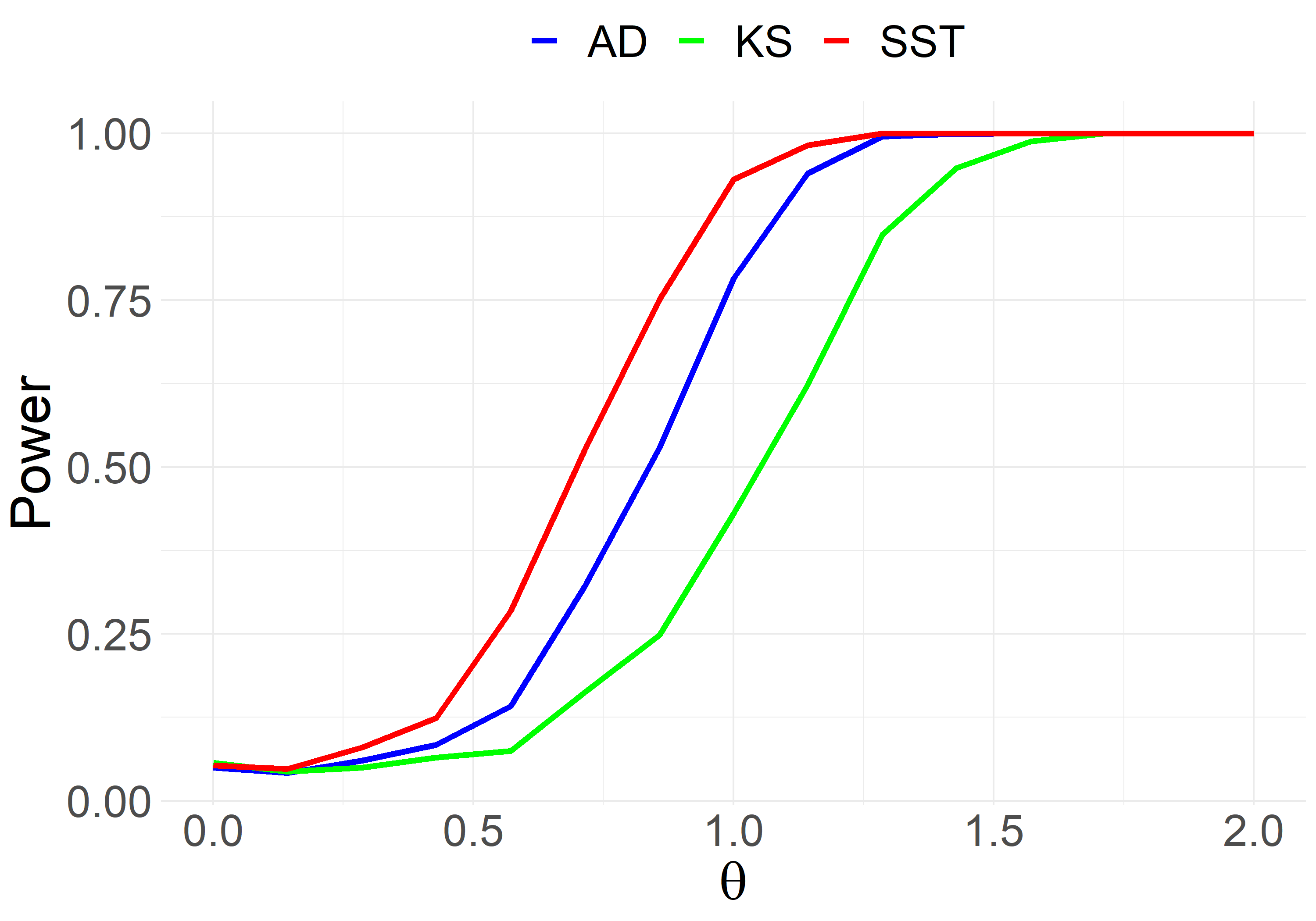}
}
\subfloat[$t(\nu=\theta^{-1})$]{\includegraphics[scale=.2]{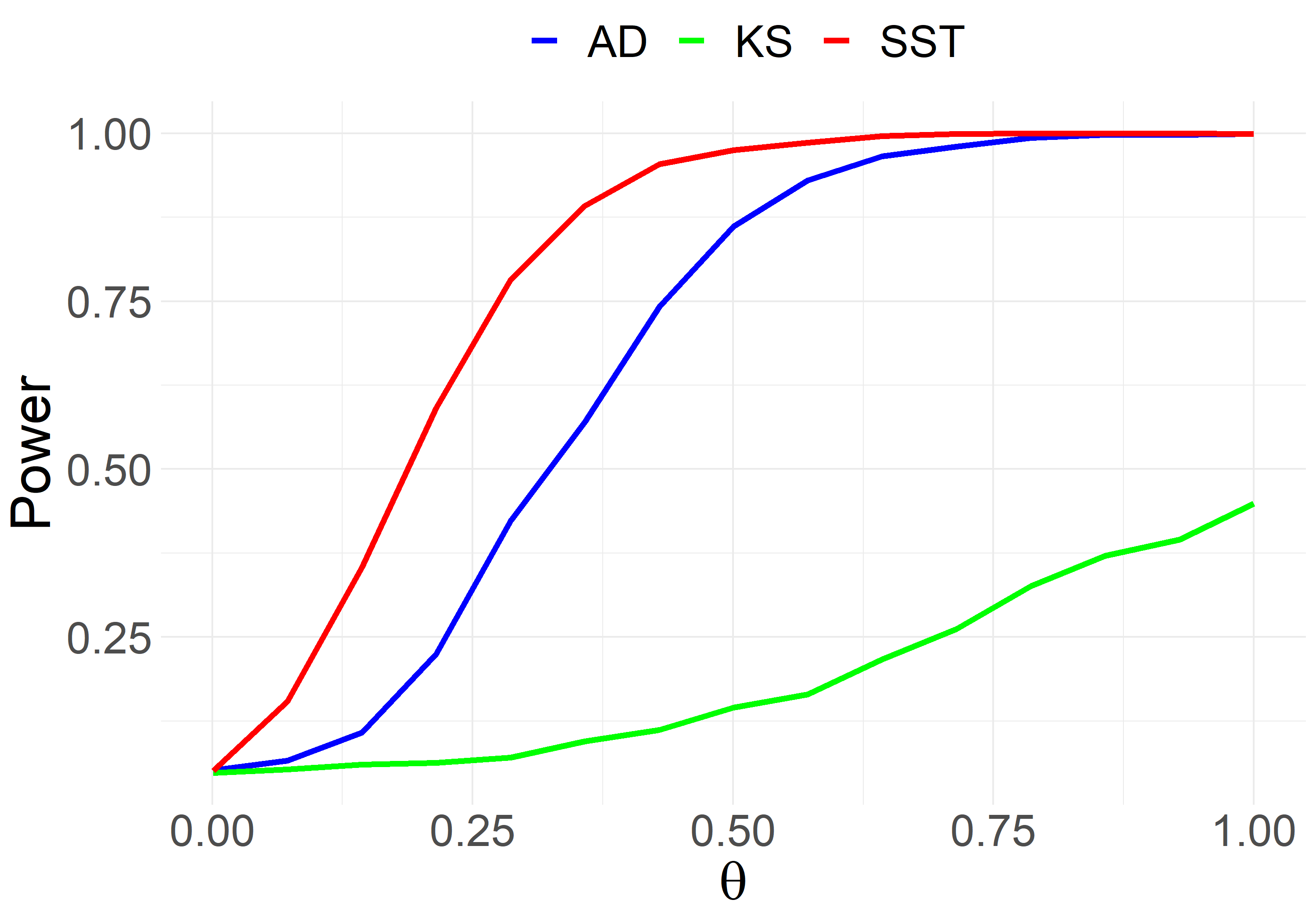}
}

\caption{Power for SST and competing methods in all univariate examples described in \cite{Ceregatti2020}. SST performs well in all scenarios, and leads to better results in four of them.} 
\label{fig:comparisons_other}
\end{figure}

\textbf{Univariate settings.} Next, we compare the performance of SST with
Anderson-Darling (AD; \citealt{anderson1952asymptotic}) and Kolmogorov–Smirnov (KS; \citealt{kolmogorov1933sulla,smirnov1948table})  in univariate settings. 
Figure \ref{fig:comparisons_other} shows that SST is competitive with the other methods in all settings, and gives  substantially better power in the four settings: gaussian with unknown variance, beta distribution, mixture of normals and t distribution.

\begin{figure}[H]
\centering
\subfloat[$\bm{\mu}(\theta)$]{\includegraphics[scale=.2]{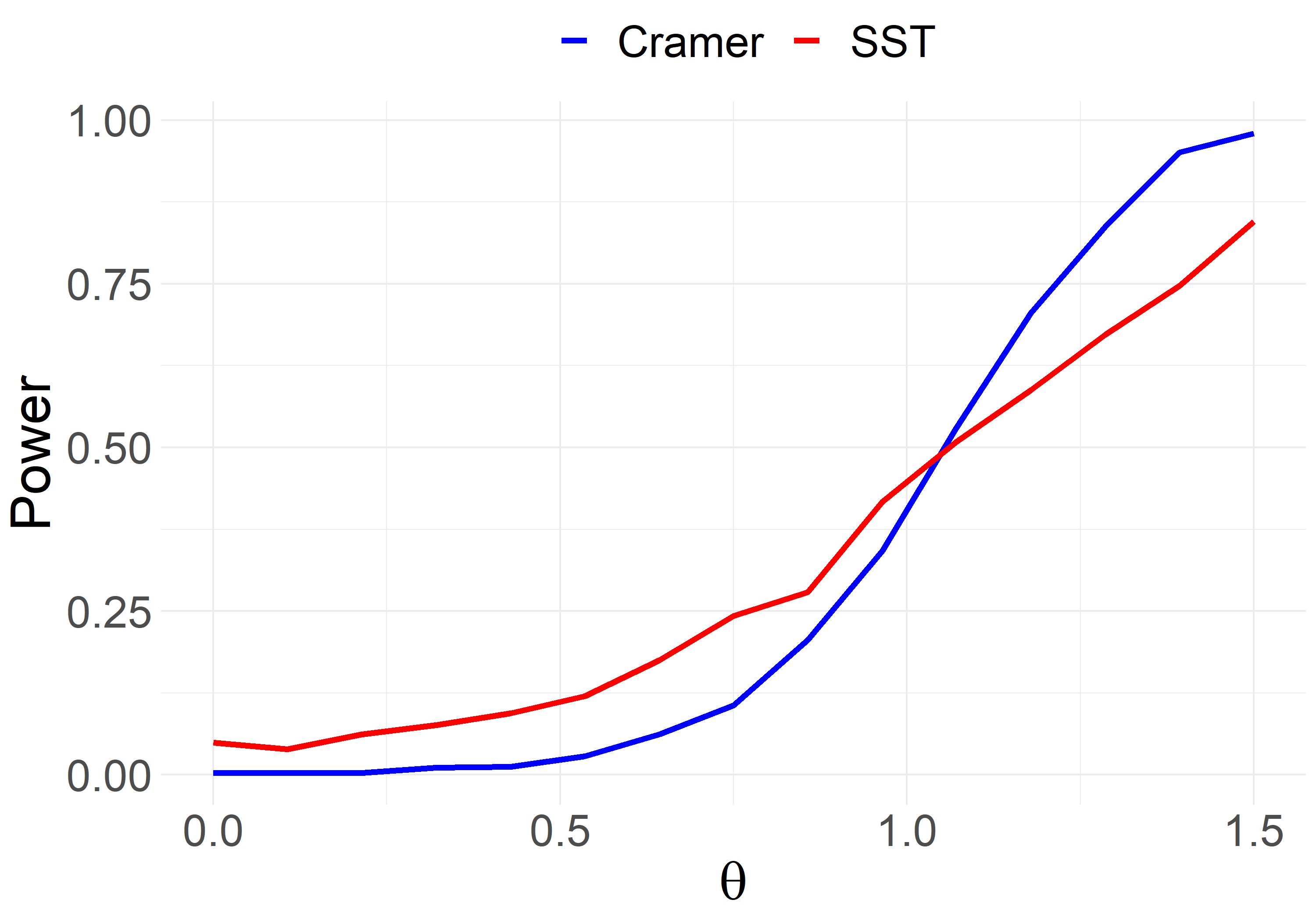}
}
\subfloat[$\Sigma(\theta)$]{\includegraphics[scale=.2]{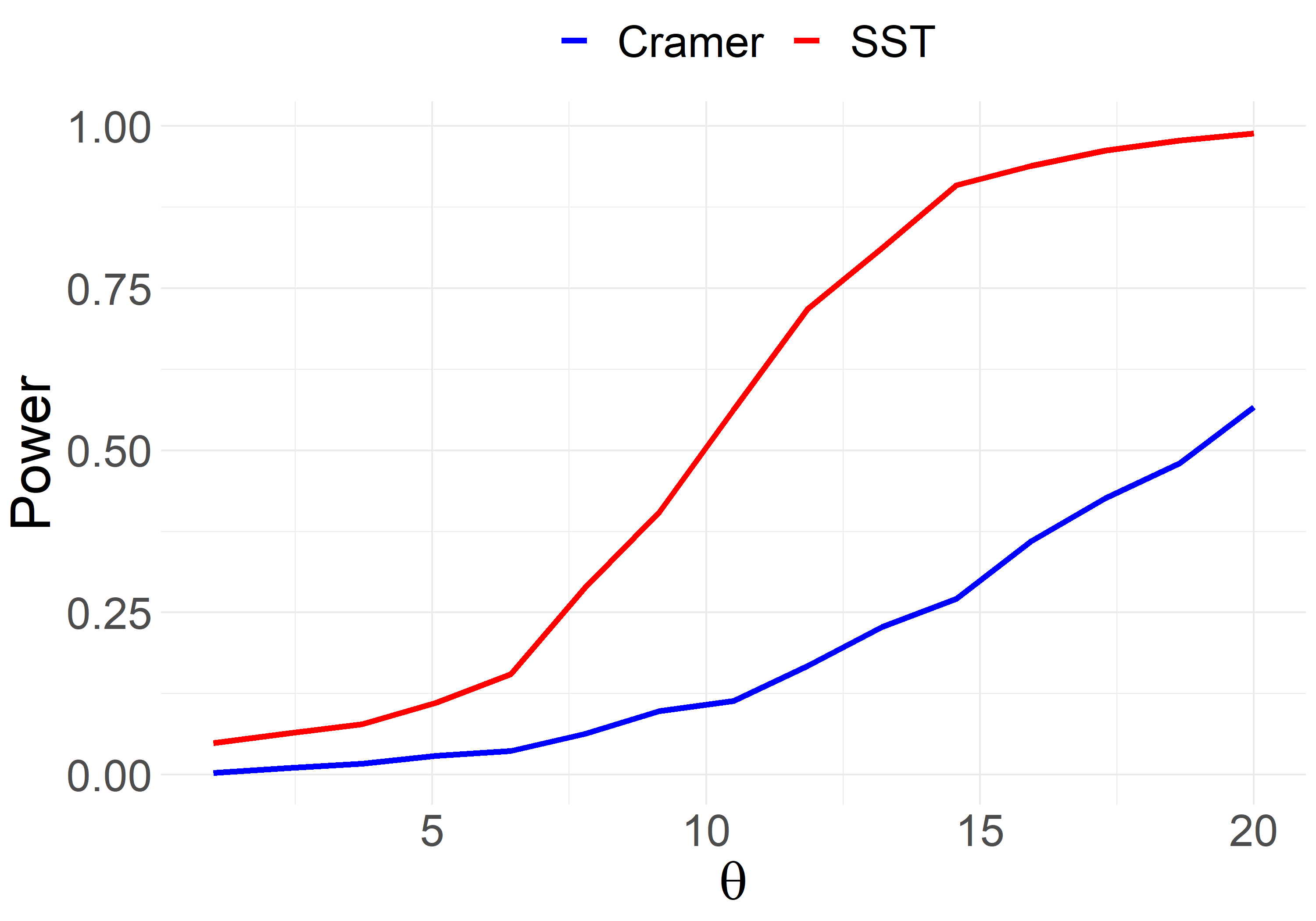}
}

\caption{Comparison in high dimension space in two situations. Standard multivariate normal distribution with dimension 100
is the null hypothesis in both of them, with variation in the mean vector $\bm{\mu}$ in (a) and in covariance matrix $\Sigma$ in (b). Our test performs well alongside Cramer.} 
\label{fig:multnorm}
\end{figure}

\textbf{Multivariate settings.}
Next, we compare the power of our method and the Multivariate Nonparametric Cramer test (Cramer; \citealt{baringhaus2004new}) in high dimension cases.

\begin{enumerate}[label=(\alph*)]
    \item Multivariate-Normal $\mu$ shift: $F_0 \sim \text{Multivariate-Normal}_{100}((0,\cdots,0),\mathbb{I}_{100})$ and $\textbf{X} \sim \text{Multivariate-Normal}_{100}((\theta,0,\cdots,0),\mathbb{I}_{100})$, $\theta = 0, \cdots, 1.5$.
    \vspace{3mm}
    
\item Multivariate-Normal $\Sigma$ shift: $F_0 \sim \text{Multivariate-Normal}_{100}((0,\cdots,0),\mathbb{I}_{100})$ and $\textbf{X} \sim \text{Multivariate-Normal}_{100}((0,\cdots,0),\left[ \begin{smallmatrix} 
\theta&0  \\ 
0&\mathbb{I}_{99} \\
\end{smallmatrix} \right])$, $\theta = 1, \cdots, 20$.
\end{enumerate}

In both scenarios the null
distribution is a multivariate normal distribution with dimension 100, zero mean vector 
and identity matrix as the covariance matrix. In the first the deviation of null distribution occurs in the mean of just one
component, this mean is $\theta$.
In second the deviation occurs in the variance of just one
component, this variance in this case is $\theta$.

While our approach is competitive against Cramer in the mean vector change setting, it clearly outperforms that test statistic in the covariance matrix change setting.
Our test is competitive against the test of Cramer \citep{baringhaus2004new}  as shown in the Figure \ref{fig:multnorm}. 

\section{Application to MNIST dataset}
\label{sec:app}

In this section we apply SST to the MNIST dataset \citep{deng2012mnist} and show that how our test can be used to interpret the difference between the null hypothesis and the true distribution of the data.

MNIST is a dataset composed by handwritten digits of two groups, students and employees. It is a modification of NIST (\url{https://www.nist.gov/srd/nist-special-database-19}) that attempted to minimize the differences  between train and test data observed in that dataset.

First, we applied our test to detect potential differences between the distribution of train and test dataset. That is, the null hypothesis establishes that the train and test data have the same distribution over the images. We sample from $F_0$ by sampling with replacement from the train set.
Our test yields a
p-value smaller than 0.001, and thus we conclude there exists a difference between the training and testing data.  Next, for each $d \in \{0,1,\ldots,9\}$, we test: 
$$H^d_0: F_{\text{train}}(\x|d)=F_{\text{test}}(\x|d),$$
where $F(\x|d)$ is the distribution over an image conditional on the digit being of type $d$. Table \ref{tab:pvalores}
shows that only digits $4$ and $5$ have p-values above $0.05$, and there exists evidence that train and test data are different for almost all digits.

\begin{table}[ht]
\label{tab:pvalores}
\caption{
P-Values of our test for all digits of the MNIST dataset. The distribution of the images from the train set is significantly different from those of the test set for most of the digits.  }
\centering
\begin{tabular}{|c|c|c|c|c|}
\hline
Digit & p-Value & \multirow{6}{*}{} & Digit & p-Value \\ \cline{1-2} \cline{4-5} 
0      & < 0.001  &                   & 5      & 0.057   \\ \cline{1-2} \cline{4-5} 
1      & 0.01    &                   & 6      & < 0.001  \\ \cline{1-2} \cline{4-5} 
2      & < 0.001  &                   & 7      & 0.001   \\ \cline{1-2} \cline{4-5} 
3      & < 0.001  &                   & 8      & < 0.001 \\ \cline{1-2} \cline{4-5} 
4      & 0.267   &                   & 9      & < 0.001  \\ \hline
\end{tabular}
\end{table}

\begin{figure}[H]
    \centering
    \includegraphics[scale=0.21]{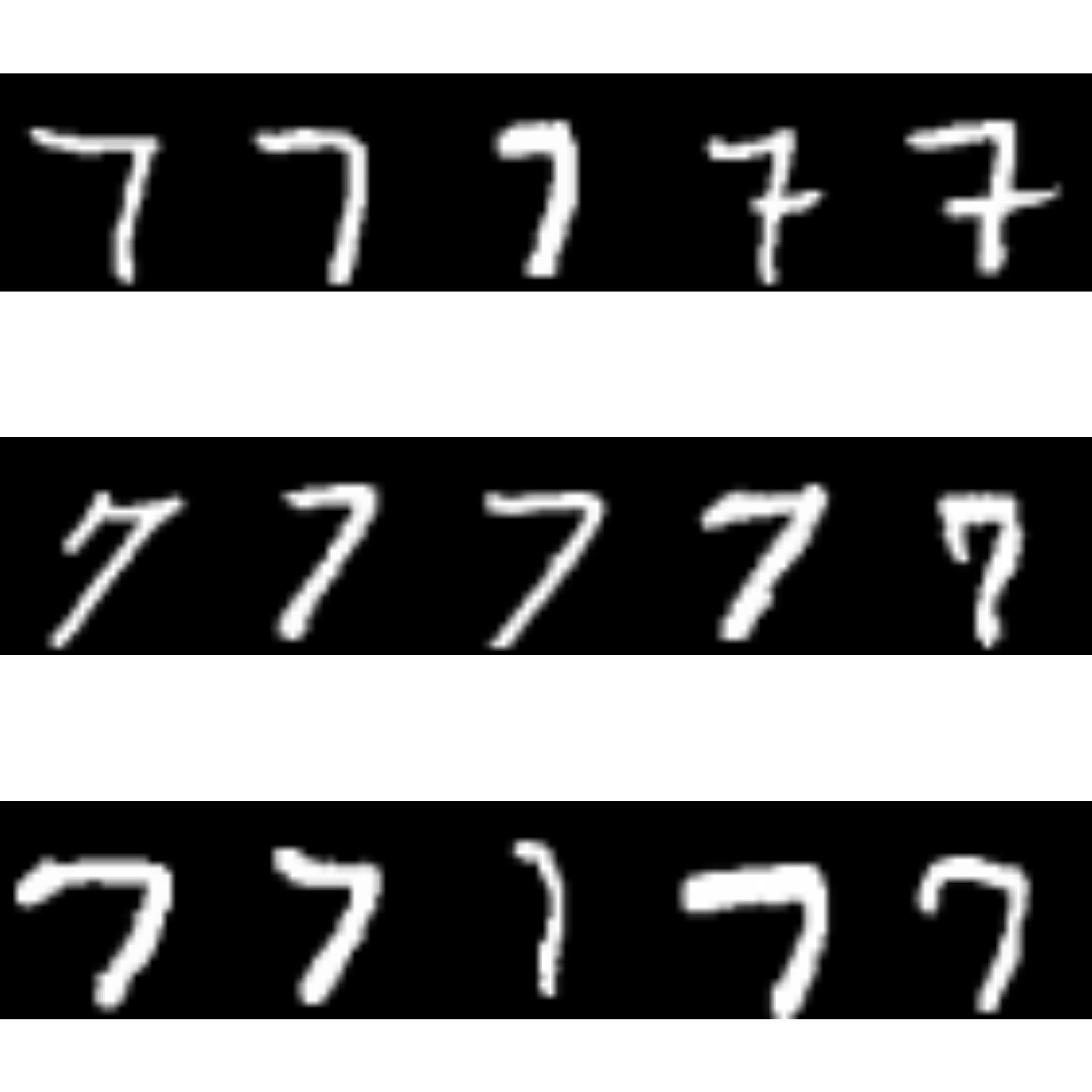}
    \caption{Pictures of number seven. Lines indicates indicates less plausibility, the same and more plausibility 
of observation of the test sub dataset against the training, respectively from top to the bottom.}
    \label{fig:mnist}
\end{figure}

Our test not only gives a reject/not-reject decision; it also offers insights about how the distributions differ.  
In order to illustrate how such insights can be obtained, we first estimate the Random-Nykodym derivative (Equation \ref{eq:radon}) using the estimated $\theta_i$'s (we use a cutoff of $I=10$ and $\epsilon=6392915$). We then rank each image in the test data according to this number.
 Values greater than 1 signify images that are more likely to be present within the test dataset, whereas values below 1 indicate a higher propensity for occurrence within the training dataset. In Figure \ref{fig:mnist}, the top row showcases images associated with the smallest Random-Nykodym derivative values, the middle row displays those with values closest to 1, and the bottom row presents images with the highest values. Notably, the images in the top row of the figure exhibit diminutive derivatives, implying their reduced likelihood of appearing within the test dataset compared to the training dataset. This observation is particularly evident in the context of crossed sevens.

\section{Final remarks}
\label{sec:final}

We have shown how smooth tests can be used on high-dimensional data. The key idea is to use spectral bases, which automatically adapt to the geometry of feature space. 
We also created a version of our test that is robust to the selection of tuning parameters.

We have shown evidence that SST has high power, and also that it offers useful insights on how the distributions being tested differ. Moreover, it naturally yields an estimate of the density of the data if the null hypothesis is rejected.
This paper therefore solidifies the standing of smooth tests as a versatile tool in the analysis of high-dimensional data.

\section{Acknowledgments}

This study was financed in part by the Coordenação de Aperfeiçoamento de Pessoal de Nível Superior - Brasil (CAPES) - Finance Code 001.
  Rafael Izbicki is grateful for the financial support of FAPESP (grant 2019/11321-9) and CNPq (grants 309607/2020-5 and 422705/2021-7).

\appendix

\section{Algorithms}
\label{sec:algorithms}

\begin{algorithm}[!ht]
	\caption{
	Estimate eigenfunctions and eigen values for a pair of bandwidth and threshold indexed by $\lambda$ 
	}\label{alg:eigenf}
	\algorithmicrequire \ {\small Cutoff and kernel hyperparameters, for example, bandwidth
}\\
	\algorithmicensure \ {\small Estimated eigenfunctions and eigenvalues for $\lambda$}
	\begin{algorithmic}[1]
	
	    \State Simulate $C_0$ a sample of $H_0$  and size $m$ to estimate the basis.
	
    	\State Calculate the matrix of kernels $\mathbb{K}$

	    \State Get right eigenvectors of $\mathbb{A}$ with norm equals to $\sqrt{n}$
	    
        \State \textbf{return} $\{\{\lambda_i\}_{i=0}^I,\{\psi_i(\y_j)\}_{j=1}^m\}_{i=0}^I$
	\end{algorithmic}
\end{algorithm}

\bibliographystyle{spbasic}      
\bibliography{references.bib}

\end{document}